\documentclass{iau}
\usepackage{graphicx,natbib}
 
\newcommand{\apj}{ApJ}           
\newcommand{\mnras}{MNRAS}       

\newcommand{\aap}{A\&A}

\newcommand{\apjs}{ApJS}           

\title{CLASH-VLT: Is there a dependence in metallicity evolution on galaxy structures?}

\author[Kuchner]{Kuchner, U.$^1$, Maier, C$^1$, Ziegler, B.$^1$, Verdugo, M$^1$, Czoske, O.$^1$, Rosati, P.$^2$, Balestra, I$^3$, Mercurio, A., Nonino, M., CLASH team}

\affiliation{$^1$University of Vienna, Department of Astrophysics,\\ T\"urkenschanzstr. 17, 1180 Vienna, Austria \\
email: {\tt ulrike.kuchner@univie.ac.at}\\
$^2$University of Ferrara, Department of Physics and Earth Science\\
$^3$INAF, Osservatorio Astronomico di Trieste\\
}

\pubyear{2014}
\volume{309}
\jname{Galaxies in 3D across the Universe}
\editors{B. L. Ziegler, F. Combes, H. Dannerbauer, M. Verdugo, eds.}

\begin{document}

\maketitle

\begin{abstract}  
We investigate the environmental dependence of the mass-metallicty (MZ) relation and it's connection to galaxy stellar structures and morphologies.
In our studies, we analyze galaxies in massive clusters at z$\sim$0.4 from the CLASH (HST) and CLASH-VLT surveys and measure their gas metallicities, star-formation rates, stellar structures and morphologies. We establish the MZ relation for 90 cluster and 40 field galaxies finding a shift of $\sim$-0.3 dex in comparison to the local trends seen in SDSS for the majority of galaxies with $logM<10.5$.
We do not find significant differences of the distribution of 4 distinct morphological types that we introduce by our classification scheme (smooth, disc-like, peculiar, compact). 
Some variations between cluster and field galaxies in the MZ relation are visible at the high mass end. However, obvious trends for cluster specific interactions (enhancements or quenching of SFRs) are missing. In particular, galaxies with peculiar stellar structures that hold signs for galaxy interactions, are distributed in a similar way as disc-like galaxies - in SFRs, masses and O/H abundances.
We further  show that our sample falls around an extrapolation of the star-forming main sequence (the SFR-M$_{\star}$ relation) at this redshift, indicating that emission-line selected samples do not have preferentially high star-formation rates (SFRs). However, we find that half of the high mass cluster members ($M_{\star}>10^{10}M_{\odot}$) lie below the main sequence which corresponds to the higher mass objects that reach solar abundances in the MZ diagram. 
\end{abstract}

\firstsection
\section{Determination of morphologies and gas abundances}
Various studies of the relation between stellar masses and metallicities of star-forming galaxies have outlined their importance for the understanding of cosmic evolution of star-formation rates and galaxy properties. In our investigation we tie in further information of galaxy structures and morphologies across different environments at z$\sim$0.4. 
Utilizing HST data from the CLASH program (Postman et al. 2012), 
we analyze galaxies in the field of 3 very massive clusters MACS1206, MACS0329 and MACS0416, complemented by Subaru BVRIz imaging which maps the large scale environment of the clusters. We use CLASH-VLT VIMOS spectra over the whole area of 30$\times$30 sq.arcmin to measure accurate star-formation rates and oxygen abundances (gas metallicities) with 5 strong emission lines [OII], Hbeta, [OIII], Halpha and [NII] (see Maier et al. 2014). 
To quantify the stellar structure, we fit the 2D light distribution of galaxies in Subaru images applying the newly developed Galapagos2 algorithms (H{\"a}u{\ss}ler et al., 2013) of the MegaMorph project 
we calculate robust structural parameters over all available Subaru bands simultaneously down to masses of $logM\sim$8.5 of 90 cluster and 40 field galaxies. 
To analyze any morphological dependences, we classify the galaxies according to four distinct types. For this, we expand a decision tree (e.g. Nantais et al. 2013) to our needs and include the new capabilities of MegaMorph. In short, (B/T)-measurements from 1- and 2-component light profile fittings with varying sets of bulge and disc profiles, as well as color information and visual classifications lead to 4 distinct classes: compact, peculiar, disc-like and objects with smooth, regular structures.

\section{Environmental and morphological dependance of the MZ relation}
Gas metallicities of galaxies depend on their stellar mass: studies of the mass-metallicity (MZ) relation across a range of redshifts show that galaxies with higher masses form a plateau of higher (solar-like) abundances, whereas towards lower masses a negative slope of the MZ relation is seen. 
Both the zero point and slope evolve with redshift (Mannucci et al. 2009, Maier et al. 2014).
However, our project is the first to investigate environmental dependencies in very massive clusters ($>10^{15} M_{\odot}$). For z$\sim$0.4 galaxies, we measure an average evolution of the MZ relation of a factor of 2 but see a shift of the mass threshold below which one finds low metallicities to lower masses for cluster environments.
We also see that at z$\sim$0.4, galaxies of different types are similarly distributed in the MZ plot, however subtle differences are visible for higher stellar masses and bigger sizes. This region is predominantly occupied by the disc-like star-forming population of our sample, both in the field and cluster. 
Differences due to environmental effects are more obvious: 
On average, while the most massive field galaxies of our sample follow the local relation reported by Tremonti et al. 2004, a few cluster members exceed solar abundances. This is independent of their morphological type.
These are the same galaxies we see below the Main Sequence in the sSFR-Mass diagram. Since peculiar galaxies are found at both high and low metallicities, we can speculate that the time scales for any environmental effects on star formation activity must be short.

\begin{figure}
\centering
\includegraphics[width=\columnwidth]{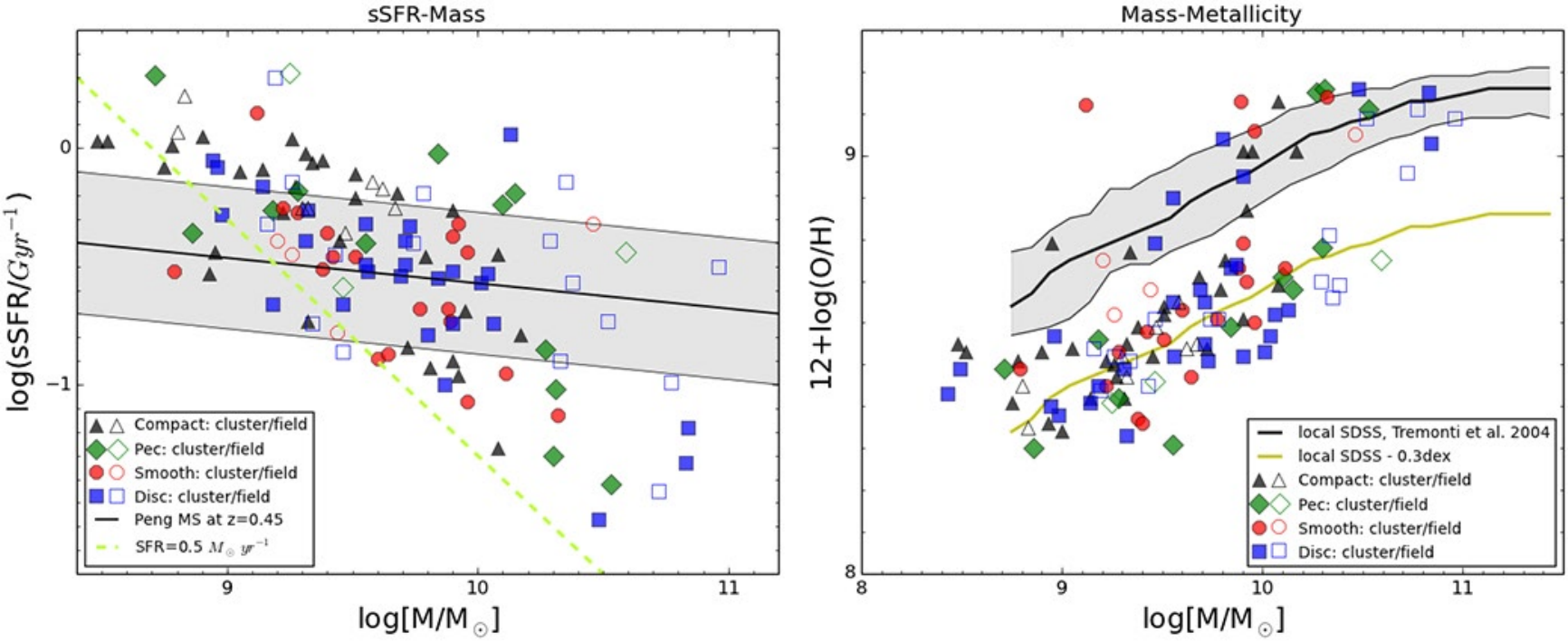} 
\caption{Left: sSFR-mass relation of 130 sample galaxies including morphological information. Right: MZ relation of the same compared to local SDSS measurements.
}\label{fig:fig1}
\end{figure}


\noindent

\end{document}